
%
%
%
%
%
%
%
\magnification 1200
\baselineskip=20.10pt
\def\boxx{\mathop{\hbox to 0pt{$\sqcup$\hss}\hbox to
0pt{$\sqcap$\hss}\phantom\nabla}}
\def\da{\dot a}
\def\na{\nabla}
\def\dph{\dot\phi}
\def\ddph{\ddot\phi}
\def\dda{\ddot a}
\def\ap{\alpha^{\prime}}
\def\b{\beta}
\def\s{\sigma}
\def\e{\epsilon}
\def\tg{\tilde g}
\def\tr{\tilde R}
\def\tess{\tilde s}
\vskip 2.5in
\rightline{hep-th/9308023}
\def\cl{\centerline}
\centerline{\bf String-Dominated Cosmology}
\vskip 1in
\cl{Dalia S. Goldwirth\footnote*{dalia@cfa.bitnet}}
\cl{Harvard-Smithsonian Center for Astrophysics}
\cl{60 Garden St.}
\cl{Cambridge, MA  02138}
\cl{USA}
\vskip .5in
\cl{Malcolm J. Perry\footnote{$^{\dagger}$}{malcolm@amtp.cam.ac.uk}}
\cl{DAMTP}
\cl{University of Cambridge}
\cl{Silver St.}
\cl{Cambridge CB3 9EW}
\cl{ENGLAND}

{\bf Abstract:}

If string theory controls physics at the string scale, the dynamics of
the early universe before the GUT era will be governed by the low-energy
string equations of motion. Studying these equations, we find that depending
on the initial conditions when the stringy era starts, and on the time when
this era ends, there are a variety of qualitatively distinct types of
evolution.
Among these is the possibility that the universe underwent
a period of inflation. A by-product of this analysis is the observation
that it is often possible to erase any evidence of a dilaton at late times.

\par\vfil\eject

In this paper we discuss the solutions of the lowest order string beta-function
equations that represent homogeneous isotropic four-dimensional spacetimes.
The basic ingredient of this calculation is therefore the massless states
of the string,
and for simplicity, we consider only the closed bosonic string. It is
quite straightforward to extend our technique to other types of string theory.
This type of solution is of importance for a number of reasons. The first is
that it illustrates some of the possible solutions to the string equations of
motion as a problem in its own right. One might think that the only solution
to the bosonic string equations that is consistent with having spatial
sections of spacetime being surfaces of constant curvature would be the well
known example of 26-dimensional Minkowski spacetime. However, it is
possible to trade off some of these extra dimensions in return for spacetime
curvature, and this is precisely what we do here.

Another reason for studying the problem is  physically motivated.
If string theory really controls physics at the string scale (presumed to be
roughly Planckian although it could be at a significally lower scale),
then from this era down to whenever the stringy
symmetries are broken to yield the physics of (presumably) the GUT era
universe, it seems reasonable to suppose that the dynamics of the
universe will be governed by the low-energy string equations of motion.
These are the equations to be studied in the remainder of this paper.

Similar, but much more restricted results have previously been obtained by
Antoniadis, Bachas, Ellis and Nanopoulos [1], Love and Bailin [2],
Campbell, Kaloper and Olive [3], Tseytlin and Vafa [4], and Tseytlin [5].
A number of other related references can be found in [5].
Their
solutions are all special cases of our results which in some sense are a
complete set of solutions that are consistent with symmetries of the spacetime
that we are imposing.

In the closed bosonic string theory, the long range massless fields
are the dilaton $\Phi$, the axion field
strength $H_{abc}=H_{[abc]}=6\partial_{[a}B_{bc]}$, which is derived from the
two-form potential $B_{ab}$,  and the graviton, or equivalently, the
spacetime metric tensor $g_{ab}$.  Condensates of these fields can be treated
in
a way consistent with the symmetries of the string and then obey the
$\beta$ -function equations [6].  If we work in the string frame,
these equations
can all be derived from the action
$$
I=\int_M d^4x g^{1/2}e^{\phi}\left(c-R-(\na\phi)^2+{1\over
12}H_{abc}H^{abc}\right)\eqno(1)
$$
where $R$ is the Ricci scalar of the metric $g_{ab}$, $c$ is a constant,
related
to the central charge of the string theory, and the integral is taken over all
of spacetime $M$. The scale of string physics is determined by by
$\alpha^\prime,$
the inverse string tension. Provided that the spacetime curvature is small on
the
string scale, then equation (1) is a complete description of the massless modes
of the
string. However,
for  curvatures large on the scale of string physics,
this action needs to be modified by terms in $\alpha^\prime$; however  in the
era
we are considering, they are negligible.

Variation of this action with respect to the metric yields the string analog of
Einstein's equation
$$
R_{ab}=\na_a\na_b\phi+{1\over 4}H_{acd}H_b^{cd}\ \ \ . \eqno(2)
$$
Variation with respect to $B_{ab}$ gives the axionic analog of Maxwell's
equation
$$
\na_aH^{abc}+\na_a\phi H^{abc}=0\ \ \ . \eqno(3)
$$
Finally, variation of the dilaton gives rise to
$$
c=R-(\na\phi)^2-2\boxx\phi+{1\over 12}H_{abc}H^{abc}\ \ \ . \eqno(4)
$$
The constant $c$ is given by $D-26\over 3\ap$ for the bosonic string, and
${3/2D-15\over 3\ap}$ for the heterotic or superstring, where $D$ is the
dimensionality of spacetime and $\ap$ is the inverse string tension.
However, $c$ can be changed from these values by coupling some conformal
field theory to the string, and so $c$ can have any value in practice.  One
often studies critical string theory where $c=0$.  The reason for such a
choice is that it enables Minkowski spacetime to be a stable ground state
for the string.  However, we are interested in cosmological solutions in
string theory, and so this is no longer a meaningful restriction.  We
regard $c$ as an arbitrary constant in what follows.

The $\b$-function equations (1-4) describe physics as seen from the
viewpoint of the string.  However, they are not convenient for
understanding gravitational phenomena, because the coefficient of the Ricci
scalar depends on the dilaton field.  An easy way to look at spacetime
physics is to perform a conformal transformation on the metric so as to
eliminate the dilaton-dependent term.  This conformal frame is usually
referred to as the Einstein frame, whereas the original one is termed the
string frame.  Such a conformal transformation yields a new metric $\tg_{ab}$
given by
$$
\tg_{ab}=e^{-\phi}g_{ab}\eqno(5)
$$
so that the action becomes
$$
I=\int d^4x \tg^{1/2}\left(ce^{-\phi}-\tr-{1\over 2}(\na\phi)^2+{1\over
12}e^{2\phi}H_{abc}H^{abc}\right)
\ \ . \eqno(6)
$$
We see from the new action that in the Einstein frame, gravitation is
described by the minimal gravitational action but matter fields couple to
gravity via the dilaton with various conformal weights.  This is
reminiscent of the situation in Brans-Dicke type of theories.  Variation of
(6) leads to the field equations in the Einstein frame.  The $\phi$
equation of motion is
$$
\boxx\phi=ce^{-\phi}-{1\over 6}e^{2\phi}H_{abc}H^{abc}\ \ \ .\eqno(7)
$$
The H equations of motion are
$$
\na_{[a}H_{bcd]}=0\ \ \ ,\eqno(8a)
$$
$$
\na_aH^{abc}-2\na_a\phi H^{abc}=0\ \ \ ,\eqno(8b)
$$
and the Einstein equation is
$$
\eqalign{R_{ab}-{1\over 2}Rg_{ab}&={1\over
2}\left(\na_a\phi\na_b\phi-{1\over 2}g_{ab}(\na\phi)^2\right)\cr
&+{1\over 4} e^{2\phi}\left(H_{acd}H_b^{cd}-{1\over
6}g_{ab}H_{cde}H^{cde}\right)\cr
&+{1\over 2}cg_{ab}e^{-\phi}\cr}\ \ \ .\eqno(9)
$$

Thus the Einstein tensor couples to the minimal energy-momentum tensor of
the dilaton, and the minimal energy-momentum tensor of the H-field
weighted by $e^{2\phi}$. The central charge term $c$ appears in a way
reminiscent of the cosmological constant, but again with a weight factor now of
$e^{-\phi}$.
One might worry about the consistency of this set of
equations.  On general grounds, one knows that (4) follows from (2) and (3)
as a consequence of the Bianchi identities.  In a similar way, one can
derive (7) as a first integral from (8) and (9).

As is apparent from (9), the right-hand side looks like a conventional
gravitational theory, and therefore one should not be surprised to discover
singularities in the solutions of (9) where the curvature blows up.  Such
singularities are however, unlike general relativity, not necessarily
unphysical.  As far as string physics is concerned, one only needs to ask
if quantities in the string frame blow up, since only then will the string
be badly behaved.  In other words, the types of singularity predicted by
the singularity theories in general relativity do not necessarily cause
breakdown of physics in string theory.

The Universe on very large scales looks like a four-dimensional
Friedman-Robertson-Walker
spacetime with a value of $k$ which cannot be observationally determined.
Therefore on large scales the spacetime metric is
$$
ds^2=-dt^2+a^2(t)d\s^2_k\eqno(10)
$$
where $d\s^2_k$ is the line element on a unit three-sphere, Euclidean
3-space, or the unit hyperboloid depending on whether $k=1,~0$, or $-1$
respectively, and $a(t)$ is the cosmological scale factor.  Consistent with
the assumption of homogeneity and isotropy implicit in this metric, we
choose the dilaton field to be independent of the special coordinates, and
the axion  field $H_{abc}=f(t)\e_{abc}$ where $\e_{abc}$ is the volume
form on the surfaces of $t=const.$  Then the requirement that $H_{abc}$ be
closed immediately implies that $f(t)$ is a constant, $f$.
We can then reduce the beta-function
equations down to a set of three coupled ordinary differential equations,
$$
c=6{\da\over a}\dph+6\left({\da\over a}\right)^2+\dph^2-{2f^2\over
a^6}+{6k\over a^2}\ \ \ ,\eqno(11)
$$
$$
\ddph=-3{\dda\over a}=6\left({\da\over a}\right)^2+{3\da\dph\over
a}-{6f^2\over a^6}+{6k\over a^2}\eqno(12)
$$
where $\dot{}=d/dt$.
We now aim to characterize all the solutions of (11) and (12).

Although (11) and (12) are the equations in the string frame, it is
straightforward to translate the solutions into the Einstein frame explicitly.
In the Einstein frame, the metric is
$$
d\tess^2=e^{-\phi}ds^2=-d\tau^2+b^2(\tau)d\s^2\eqno(13)
$$
where
$$
\tau=\int dt e^{-{1\over 2}\phi}\eqno(14a)
$$
and
$$
b(\tau)=e^{-{1\over 2}\phi(\tau)}a(\tau)\ \ . \eqno(14b)
$$

To simplify the discussion of these equations, firstly we consider the
special case of the critical string $(c=0)$ and a flat universe $(k=0)$.  We
can now eliminate $f$ and find that
$$
\ddot\phi=-{3\dda\over a}=-12\left({\dot a\over a}\right)^2-15{\dot a\over
a}\dph-3\dph^2\ \ \ .\eqno(15)
$$
Perhaps the easiest way to find all the solutions to these equations is to
change variables into
$$
\eqalign{p&=\dph\cr \chi&=\da/a\cr}\eqno(16)
$$
yielding
$$
\eqalign{\dot p&=-12\chi^2-15p\chi-3p^2\cr
\dot \chi&=3\chi^2+5p\chi+p^2\cr}\ \ \ .\eqno(17)
$$

There are several simple analytic  solutions that can easily [1,5] be found
from (17).  They
are
$$
p={A\over t-t_0},~~~\chi={B\over t-t_0}\eqno(18)
$$
with any of
$$
\eqalign{{\rm i)} A&=0~~~~~~~~~B=0\cr
{\rm ii)} A&=-2/3~~~~~B=1/3\cr
{\rm iii)} A&=1-\sqrt{3}~~~B=1/\sqrt{3}\cr
{\rm iv)} A&=1+\sqrt{3}~~~B=-1\sqrt{3}\cr}\eqno(19)
$$
Case (i) is flat Minkowski spacetime (in either the Einstein frame or the
string frame) was first discussed by
Antoniadis, Bachas, Ellis and  Nanopoulos, [1].  Case (ii) is
$a(t)=a_0t^{1/3}$ and $\phi(t)=\phi_0-2/3 lnt$ (setting $t_0=0$).
If we transform this into the
Einstein frame $(\phi_0=0)$, then
$$
\tau={3\over 4}t^{4/3}~~~~~ {\rm and} ~~~~~~ b=b_0\tau^{1/2}\ \ \ .\eqno(20)
$$
In other words, case (ii) is identical to a
radiation-dominated universe.  It is singular at $t=0$, even in the string
frame, and unphysical since one finds that $f^2<0$.
Cases (iii) and (iv), were originally found by Tseytlin, [5],
have $f^2=0$ and similar power law expansions and
contractions.  Case (iii) is $a(r)=a_0t^{1/\sqrt3}$ while case (iv) is
$a(r)=a_0t^{-1/\sqrt3}$.

The remaining solutions can be explored by examining the phase-plane
portrait using the fact that Eq. (17) results in
$$
{dp\over d\chi}=-{12\chi^2+15p\chi+3p^2\over 3\chi^2+5p\chi+p^2}\ \ \
.\eqno(21)
$$
It is shown in figure (1),  the region of physical solutions is bounded by
the exact solutions of case (iii) and (iv) which are given by
$$
p = (-3\mp \sqrt3)\chi \ \ . \eqno(22)
$$
The evolution is in the direction of the arrows indicated.
With the exception of the exact solutions (iii) and (iv), the spacetimes
are nonsingular.
There are two types of evolution depending on which side of the phase
plane the trajectories begin.
In both cases
tracing the solutions backwards in time reveals that
at earlier times they were repelled from the contracting version of either the
exact solution (iii) or (iv).
For solutions that expand rapidly at late
times, the exact solution (iv) is an attractor. These type of solutions
occupy the left-hand half of figure (1). The solutions that occupy the
right-hand half of figure (1) are attracted to solution (iii) and on it to the
$(0, 0)$ point, i.e the universe comes to a halt asymptotically.
In Figure (2) we plot $a(t)$ and $\chi(t)$ for two typical cases.

We turn now to a more complicated situation in which the spatial
sectors of the universe are not flat.
In order to get a phase-plane picture similar to the one we got in the flat
case
, we change variables from $t$ to $\theta$ such that
$$
{d\theta \over dt} = {1 \over a}\ \ .  \eqno(23)
$$
Using $\theta$ as a dynamical variable we now define $\chi$ and $P$ to be:
$$
\chi = {a^{\prime}\over a} \ \ \ \ \ \ \  P = \phi^{\prime}  \ \ , \eqno (24)
$$
where $'\ \ = d/d\theta$. The analog of Equation (17) can then be rewritten as
$$
P^{\prime} = -14\chi P -12\chi^2 -12k -3P^2 \eqno (25)
$$
$$
\chi^{\prime} = 5\chi P +4\chi^2 + 4k + P^2 \ \ .
$$
To check consistency, the solutions of Eqs. (25) must
satisfy Eq. (11), which in terms of the new variables takes the form
$$
6\chi P + 6\chi^2 + P^2 + 6k = 2f^2/a^4  \ \ . \eqno (26)
$$
We examine now the phase-plane properties derived from Eq. (25). Firstly we
construct $dP/d\chi$ and find that
$$
{dP \over d\chi} = -3 + {\chi P \over 5\chi P + 4\chi^2 + 4k + P^2} \ \ .
\eqno (27)
$$
The analytic solutions which bound the regions of the physical solutions
are given by the requirement that $f^2 = 0$.
{}From Eq. (26) it follows that
$$
6\chi P + 6\chi^2 + P^2 + 6k = 0 \ \ . \eqno (28)
$$
Therefore, in the asymptotic regime where $\chi \gg k$, the solutions which
bound the physical solutions are given as
before by Eq. (22) for the new $P$ and $\chi$.
However, the exact solution to Eq. (28) is
$$
P = -3\chi \pm \sqrt{3\chi^2-6k} \ \ . \eqno (29)
$$
Those, there is radically different behavior in the case $k=1$ compared to
$k=-1$ when one gets close to the origin of the $\chi$-$P$ plane.
We denote by (ASiii) and (ASiv) those lines that are asymptotic to solution
(iii) and (iv),
and turn now to examine the properties of the trajectories in a closed
universe.

For $k=1$, physically valid solutions penetrate into the
region that was forbidden in the case that $k=0$.
As in the flat case, there are two types of trajectories
depending on which part of the phase-plane shown in Fig. 3 they occupy.
All solutions are asymptotic to the lines given by Eq. (28).
 However, all the trajectories end
up on the expanding part of (ASiv). These on the upper part of the
$(\chi, P)$ plane are repelled from the upper part of (ASiv), and
are attracted to the collapsing part of (ASiii). The collapse slows
down as the trajectories approach solution (ASiii).
The trajectories route depend on whether they start below or above solution
(iv).  Trajectories that start from below go through the left-hand part
of the phase plane in figure (3) and continue to collapse with different rate.
The ones that start from above go through the right-hand part of the phase
plane. In this case the collapse turns into a slow expansion along the
expansion
part of solution (ASiii). In both cases
the further they started up on (ASiv) the further they will
go along solution (ASiii).
In both cases the collapse or the slow expansion turn into rapid expansion as
the trajectories are repelled from (ASiii) towards the expanding part
 of solution (iv).
In figure 4 we show two typical sets of evolution for $a$, the scale factor,
and $\chi$ to demonstrate each of these histories. One can see that the time
scale on
which changes take place is completely different but eventually both reach the
point of fast expansion. In both cases the scale factor becomes infinite in
a finite time.

Now, we turn our attention to an open universe by setting $k=-1$.
There is an elementary exact solution with $P=0$ and $\chi = 1$
 (found already by
Tseytlin) for which
$$
a = t + t_0  \ \ \ \ \ \ \ \ \ \ \phi = \phi_0  \ \ . \eqno (30)
$$
This solution has the same form in the Einstein frame.
All the trajectories in the right-hand of figure (5) approach this solution
marked by a circle on this figure.
Tracing these trajectories to the past reveals that
all the trajectories are repelled from the contracting part of the
asymptotic solution (ASiv). Here since $f^2 = 0$ requires $6\chi P + 6\chi^2
+P^2 - 6 =0$, all the trajectories start above the exact solution (iv).
Because of this, trajectories that occupy the left-hand of figure (5) are
repelled from below by the contracting part of (ASiv). These trajectories are
attracted by the expanding part of (ASiii), i.e these universes end up in
rapid expansion. We see that in this case, unlike the case when $k=1$, there
are two completely different types of behavior depending on the initial
conditions.
Figure (6) shows two of the line evolutions of typical example of these
two different types of trajectory.  In the first case, the Universe
ends up growing linearly while in the second case it starts from linear
contraction and which then turns into rapid expansion.

We turn now briefly to the case of $c \ne 0$. The analogs of Equation (17)
are, for $k=0$,
$$
\dot P = -12\chi^2 -15 P \chi -3P^2 -3c
$$
$$
\dot \chi = 3\chi^2 +5P\chi + P^2 + c  \eqno (31)
$$
(now $\dot{}  = d /dt$ as for the $k=0$ case).
 As before the phase plane properties can be explored from
observing that
$$
{dP \over d\chi} = -3 + {\chi^2 \over 3\chi^2 + 5P\chi + P^2 + c}
\ \ . \eqno (32)
$$
The boundaries of the physical allowed regions are given by $f^2=0$, for
which
$$
6\chi^2 + 6P\chi + P^2 - c =0 \ \ . \eqno (33)
$$
The solutions to Eq. (33) are therefore
$$
P = -3\chi \pm \sqrt{3\chi^2+c} \ \ . \eqno (34)
$$
It seems therefore intuitively apparent that by comparing Eq. (34) with
Eq. (29) that $c>0$ should be qualitatively similar to $k<0$ and the effect
of $c<0$ should be qualitatively similar to $k>0$.

It is in principle possible to extend our method to deal with the cases
$k \ne 0$, $c\ne 0$ simultaneously. We do not do this here because it is
too complicated to be in good taste.
\bigskip
{\bf Conclusions}
\bigskip
 Our overall conclusions are that in some stringy era, the large scale
properties of the universe are not without some interest. Let us describe the
$c=0$, $k=0$ case first. The contracting part of the exact solutions,
the dashed line in Fig. 1, repels all the trajectories. So the general
evolution will be away from it, into the expanding phase.
As one might have expected by time-reversal invariance,
there are solutions which are the reverse of these.  Crudely speaking,
these occupy the left-hand half of figure (1) whereas the original one
occupy the right-hand half.
These solutions start from the exact solution (iii) and are repelled towards
the
exact solution (iv) and so the universe ends up in rapid expansion.
If one starts in the upper part of the diagram, the evolution will be
towards a universe that reaches constant size as $t\rightarrow\infty$, as
given by the attractor shown by the dotted line, or analytically by equation
...
If however, one starts in the lower part of the diagram, the expansion
will always lead to hyperinflation.  In our mathematical idealization, the
radius of the universe tends toward infinity in  a
finite proper time. However, it could be the case that in a realistic
theory, there is an exit regime when the effective temperature of the
universe reduces below some critical scale, presumably the GUT scale. For
$k=1,~c=0$, (or equivalently for $c<0$ and $k=0$) then for large values of
$x$ and $P$, the behavior is qualitatively similar.  However, for small $x$
and $P$, there are significant differences. If one starts in a contracted era
in
the upper part of figure 3, provided one does not start too close to the
repeller, an era of contraction will be followed by expansion, which can
be very slow for a significant time.  This is called a loitering era [7].
Eventually, the expansion will accelerate, again leading to hyperinflation
unless some exit from the string-dominated era can occur.  If one starts
very close to the repeller, then there is again a loitering phase which
occurs during the contraction.  Eventually however, expansion and then
hyperinflation will set in, again subject to the caveats that we are still
in the stringy era.  For $k=-1,~c=0$ (or equivalently $c>0$ and $k=0$), if
one starts in the upper part of the picture in a contracting phase, then this
always turns into expansion, and at late times $a(t)\sim t$.  This corresponds
to the fixed point in the $(x,P)$ plane marked by an open circle in figure 5.
Similarly, one can start at a point in the lower part of figure 5, and
again one discovers that there is an attractive hyperinflating solution.

We see therefore that during a stringy era, one can get a large variety of
different behavior depending on the initial conditions when the stringy era
sets in. Furthermore, depending on when the stringy era ends, one can find
a number of clues to how GUT era physics could have started off.

There is an addition to the cosmological interest the potential in this work
to resolve a string theory puzzle. If string theory really does describe the
physics of gravitation, it seems that there should be a massless scalar
component to the gravitational field  in the form of the dilaton. However,
as has been emphasized by Damour and Nordvedt[8], the dilaton is forced by
the cosomological expansion to the general relativistic limit. Our
results are no exception to this observation as can be seen from the fact
that $\phi$ is often driven to infinity, which corresponds to weak coupling
of the dilaton.

{\it Acknowledgements}

This work was supported in part by the Center for Astrophysics Fellowship
(DSG).
MJP would like to thank the Theoretical Astrophysics Division of the
Harvard-Smithsonian Center for Astrophysics for its hospitality while  his
work was in progress.
\par\vfil\eject
Figure Caption

\item{Fig 1.} A phase plane digram in the $(\chi,P)$ plane for $k=0$ and
$c=0$. Evolution is in the direction of the arrows. The exact solutions
$(iii)$ and $(iv)$ are identified by the half dashed half dotted lines.
The dashed part is a repeller and the dotted line is an attractor.

\item{Fig 2.} Two typical solutions of
$Log~ a$ and $\chi$ as function of $t$.

\item{Fig 3.} The same as Fig 1. for $k=1$ and $c=0$.

\item{Fig 4.} Two typical solutions of
$a$ and $\chi$ as function of $t$. The straight line corresponds to the points
marked by the circles in Fig. 3.

\item{Fig 5.} The same as Fig 1. for $k=-1$ and $c=0$.

\item{Fig 6.} Two typical solutions of
$a$ and $\chi$ as function of $t$.
\item{      a.} For trajectories that start near the origin.
\item{      b.} For trajectories that start away from the origin.

\par\vfil\eject

References:

\item {1.} I. Antoniadis, C. Bachas, J. Ellis and D.V. Nanopoulos, Phys. Lett.
B221, 393, (1988).
\item{} I. Antoniadis, C. Bachas, J. Ellis and D.V. Nanopoulos, Phys. Lett.
B257,
278, (1991).
\item{} I. Antoniadis, C. Bachas, J. Ellis and D.V. Nanopoulos, Nucl. Phys.
B328, 117, (1989).

\item{2.} D. Bailin and A. Love, Phys. Lett. B163, 135, (1985).

\item{3.} B.A. Campbell, N. kaloper and K.A. Olive, Phys. Lett. B277, 265,
(1992).

\item{4.} A.A. Tseytlin and C. Vafa, Nucl. Phys. B372, 443, (1992).

\item{5.} A.A. Tseytlin, Class. Quant. Grav. 9, 979, (1992).
\item{} A.A. Tseytlin, preprint DAMTP-92/15, hepth 9203033, (1992).

\item{6.} E.S. Fradkin, Phys. Lett. B158, 316, (1985).
\item{} C.G. Callan, D. Friedan, E, Martinec and M.J. Perry, Nucl. Phys. B262,
593, (1985).
\item{} C. Lovelace, Nucl. Phys. B273, 413, (1985).

\item{7.} V. Sahni, H. Feldman and A. Stebbins, Ap. J. 385, 1, (1992).

\item{8.} T. Damour and K. Nordtvedt, IHES preprint, IHES/P/93/16, (1993).

\bye